\documentclass[aps,prb,preprint,groupedaddress,showpacs]{revtex4-1}

\usepackage{graphicx}
\usepackage{longtable}
\begin{document}

 Manuscript prepared for Proceedings of the Combustion Institute \\
 (35th Symposium on Combustion)
\\
\\

\title{Validation of reduced kinetic models for simulations of
non-steady combustion processes}
\author{M.F. Ivanov$^1$}
\author{A.D. Kiverin$^1$}
\author{M.A. Liberman$^{2,3}$}
\email{m.liber@nordita.org}
\author{A.E Smygalina$^1$}

\medskip
\affiliation{$^1$ Joint Institute for High Temperatures,
Russian Academy of Science, Izhorskaya 13,
Bd. 2 Moscow 125412, Russia \\
$^2$ Nordita, KTH Royal Institute of Technology and Stockholm University,Roslagstullsbacken 23, 10691 Stockholm, Sweden\\
$^3$ Moscow Institute of Physics and Technology, Dolgoprudnyi, 141700, Russia}

\date{\today}
\begin{abstract}
In the present work we compare reliability of several most widely
used reduced detailed chemical kinetic schemes for hydrogen-air
and hydrogen-oxygen combustible mixtures. The validation of the
schemes includes detailed analysis of 0D and 1D calculations and
comparison with experimental databases containing data on induction
time, equilibrium temperature, composition of the combustion products,
laminar flame speed and the flame front thickness at different
pressures. 1D calculations were carried out using the full
gasdynamical system for compressible viscous thermal conductive
multicomponent mixture. The proper choice of chemical kinetics
models is essential for obtaining reliable quantitative and
qualitative insight into unsteady combustion phenomena such
as flame  acceleration and stability, ignition, transition from
deflagration-to-detonation (DDT) using a multidimensional and multiscale
numerical modeling.
\end{abstract}


\maketitle

\section{Introduction}

Even in the combustion of small hydrocarbons, the chemical
kinetics has large underlying reaction mechanisms, and for
complex hydrocarbons, the number of chemical species can be
up to several hundreds and elementary reactions up to several
thousands. Accurate knowledge of the detailed reaction mechanisms
is of great importance for understanding and a correct description
of kinetically controlled transient combustion processes such as
ignition and self-ignition processes (e.g. engine knock), flame
extinction, or transition from deflagration-to-detonation. If,
however, real three-dimensional (turbulent) flows with large
temperature and density gradients are considered, we have to
use reduced chemical schemes, since the use of detailed reaction
mechanisms involves massive computing times which can be difficult
and even impossible to implement. Development and exploitation of
reliable detailed chemical kinetic models and identification of the
important kinetic pathways and accurate kinetic-transport models
remain among the major challenges in combustion science and
technology being essential for the design of efficient and
reliable engines and for controlling emissions. The availability
of such models is essential for gaining scientific insight into
important combustion phenomena including flame acceleration and
stability, ignition processes, transition from deflagration-to-detonation
(DDT) and for the design of advanced combustion engines. Therefore,
there is great interest in simplified - reduced chemical reaction
schemes consisting of not too large number of elementary reaction.
Then the problem is to find criteria for option of a reliable detailed
chemical kinetic model. A quintessential example of chain mechanisms in
chemical kinetics and combustion science is the ${H_2}-{O_2}$ mechanism,
which has been a major topic of research for many decades.

Comprehensive numerical modeling of the unsteady combustion
should include a reliable detailed chemical reaction scheme
for the understanding of complex multiscale phenomena observed
in experiments, since a one-step model does not reproduce even
two distinct stages of the combustion reaction: induction
stage and exothermal one. It was shown in numerous numerical
studies (see e.g. \cite{MaasWarnatz88,Sloane93,LiberDDT2012,KiverIgni2013}
that in many cases both quantitative and qualitative features of
the studying processes depends on the choice of chemical kinetics
model. Moreover while studying unsteady processes such as flame
acceleration, deflagration-to-detonation transition \cite{Ivan3DDT2013}
or flames within combustors \cite{LiberNock2006,Gu2003}
one should use models correctly reproducing flame parameters
in a wide range of pressures and temperatures.

There are different reduced kinetics schemes, which validation
procedures usually use $0D$ and  $1D$ calculations and experimental
databases containing data on induction periods, equilibrium
temperature and composition of the combustion products and
laminar flame speeds. Usually, the ignition and combustion for
verification databases are studied using the experimental setups far
distinct from the setups used for transient processes studies.
In some cases it may cause considerable  differences between
experimental data and those calculated using the chosen model.
Almost all the listed parameters are defined by the thermodynamic
equilibrium laws. The only evolutionary parameter is an induction
period determining the duration of the endothermal stage.
However, in some cases the duration of the exothermal stage may
be also principal for the process definition as  the scale of
energy release determines the gasdynamical flows.

Widely accepted standard numerical procedures for ignition parameters
and for flame speed calculations are based on the $0D$ solution
and on the $1D$ solution of the eigenvalue problem, which is
distinguished mathematically from the computational gasdynamics
setup used for transient combustion simulations. However instead
of solving the eigenvalue problem one may use other approaches to
simplify full reactive gasdynamics model
(as e.g. it was done in \cite{Steph1973}).  On the other hand to
simulate a transient problem one should validate the codes
and models describing the real process involving pressure
gradients, compressibility, convection, turbulence etc.
Therefore it is important to get the solution of basic
problems using more general models which are planned to
be used in the simulation of multidimensional unsteady problems.

In the present paper we evaluate different widely used kinetic
schemes for hydrogen-air and hydrogen-oxygen combustion
\cite{OConaire2004,Konnov2004,Warnatz2006,Agafonov1998,GRI}
using the full gasdynamical models including standard transport
model \cite{Hirschfelder1964} for laminar flame characteristics.
The analysis takes into account the correlations between
evolutionary parameters (induction period and duration of
exothermal stage) and gasdynamical ones (laminar flame speed
and its thickness) for hydrogen-air and hydrogen-oxygen
mixtures at different initial pressures.

The paper is organized as follows. Section 2 is the
formulation of the problem and numerical method.
In Section 3 we perform calculations of the chemical
time scales and gasdynamical parameters of the laminar
flame in hydrogen-air mixtures. Section 4 presents analysis
of the hydrogen-oxygen flames. In Section 5 we formulate
conclusions about the applicability of the chosen kinetic
schemes in numerical simulations of the transient combustion processes.

\section{Problem setup}

The governing equations are the one-dimensional time-dependent,
multi-species reactive Navier-Stokes equations including the effects of
compressibility, molecular diffusion, thermal conduction,
viscosity and chemical kinetics for the reactive species
with subsequent chain branching, production of radicals
and energy release.

\begin{eqnarray}
\frac{{\partial \rho }}{{\partial t}} +
\frac{{\partial \left( {\rho u} \right)}}{{\partial x}} = 0,
 \label{eq1}
\end{eqnarray}
\begin{eqnarray}
\frac{{\partial {Y_i}}}{{\partial t}} +
u\frac{{\partial {Y_i}}}{{\partial x}} =
\frac{1}{\rho }\frac{\partial }{{\partial x}}
\left( {\rho {D_i}\frac{{\partial {Y_i}}}
{{\partial x}}} \right) + {\left( {\frac{{\partial {Y_i}}}
{{\partial t}}} \right)_{ch}},
 \label{eq2}
\end{eqnarray}
\begin{eqnarray}
\rho \left( {\frac{{\partial u}}{{\partial t}} +
u\frac{{\partial u}}{{\partial x}}} \right) =
- \frac{{\partial P}}{{\partial x}} +
\frac{{\partial {\sigma _{xx}}}}{{\partial x}},
 \label{eq3}
\end{eqnarray}
\begin{eqnarray}
\rho \left( {\frac{{\partial E}}{{\partial t}} +
u\frac{{\partial E}}{{\partial x}}} \right) =
 \nonumber \\
- \frac{{\partial \left( {Pu} \right)}}{{\partial x}}
 + \frac{\partial }{{\partial x}}\left( {{\sigma _{xx}}u} \right)
 + \frac{\partial }{{\partial x}}
  \left( {\kappa \left( T \right)\frac{{\partial T}}{{\partial x}}}
   \right)
    \nonumber \\ + \sum\limits_k {\frac{{{h_k}}}{{{m_k}}}
   \left( {\frac{\partial }{{\partial x}}
   \left( {\rho D_k^{}\left( T \right)\frac{{\partial {Y_k}}}
   {{\partial x}}} \right)} \right)},
 \label{eq4}
\end{eqnarray}
\begin{eqnarray}
P = {R_B}T\,n = \left( {\sum\limits_i
{\frac{{{R_B}}}{{{m_i}}}{Y_i}} } \right)\rho T
= \rho T\sum\limits_i {{R_i}{Y_i}}
 \label{eq5},
\end{eqnarray}
\begin{eqnarray}
\varepsilon  = {c_v}T + \sum\limits_k {\frac{{{h_k}{\rho _k}}}{\rho }}
= {c_v}T + \sum\limits_k {{h_k}{Y_k}},
 \label{eq6}
\end{eqnarray}
\begin{eqnarray}
{\sigma _{xx}} = \frac{4}{3}\mu
\left( {\frac{{\partial u}}{{\partial x}}} \right)
 \label{eq7}
\end{eqnarray}

Here we use the standard notations:
$P$, $\rho $ , $u$, are pressure, mass density, and flow velocity,
${Y_i} = {\rho _i}/\rho$  - the mass fractions of the species,
$E = \varepsilon  + {u^2}/2 $ - the total energy density,
$\varepsilon $ - the internal energy density,
${R_B}$ - is the universal gas constant,
${m_i}$- the molar mass of i-species,
${R_i} = {R_B}/{m_i}$, $n$ - the molar density,
$ \sigma _{ij}$- the viscous stress tensor,
${c_v} = \sum\limits_i {{c_{vi}}} {Y_i}$ - is the
constant volume specific heat,
$c_{vi}$- the constant volume specific heat of i-species,
$h_i$  - the enthalpy of formation of i-species,
$\kappa (T)$  and $\mu (T)$  are the coefficients of thermal
conductivity and viscosity, ${D_i}(T)$  - is the diffusion coefficients
of i-species,  ${\left( {\partial {Y_i}/\partial t} \right)_{ch}}$ - is
the variation of i-species concentration
(mass fraction) in chemical reactions.

The equations of state for the reactive mixture and
for the combustion products were taken with the
temperature dependence of the specific heats
and enthalpies of each species borrowed from
the JANAF tables and interpolated by the
fifth-order polynomials \cite{Warnatz2006,McBride1993}.
The viscosity and thermal conductivity coefficients
of the mixture were calculated from the gas kinetic
theory using the Lennard-Jones potential \cite{Hirschfelder1964}.

The system of gas dynamics equations is solved using
Lagrange-Euler method \cite{Belotserk1982}, which was modified
and approved by authors solving numerous 1D, 2D and 3D problems
(see e.g. \cite{LiberDDT2012,KiverIgni2013,Ivan3DDT2013,LiberNock2006}.
The system of chemical kinetics equations is solved with the aid of Gear method.

\section{Hydrogen-air mixture}

Usually the verification of the reduced kinetic schemes covers
the data on induction periods and equilibrium composition of
the products and their temperature. The later are pure thermodynamical
characteristics and are more or less in a good agreement with
experimental data. In turn these parameters together with transport
coefficients determine the laminar flame speed. To reproduce
transient processes such as e.g. flame acceleration, ignition, etc.
which are accompanied by compression and shock waves, one should
take into account parameters such as induction periods and periods
of exothermal reaction, which determine the chemical time scales
competing with transport time scales in establishing the flame
front - the zone of energy release which in fact determines the
evolution of the flame. Nevertheless, the even more important
thing is the necessity of understanding and reproducing the
pressure dependence of flame parameters. Here we present analysis
of these parameters given by the different kinetic
schemes for ${H_2}/{O_2}$ and ${H_2}$-air mixtures.

\begin{figure}
\vspace*{1mm} \centering
\includegraphics[width=9cm]{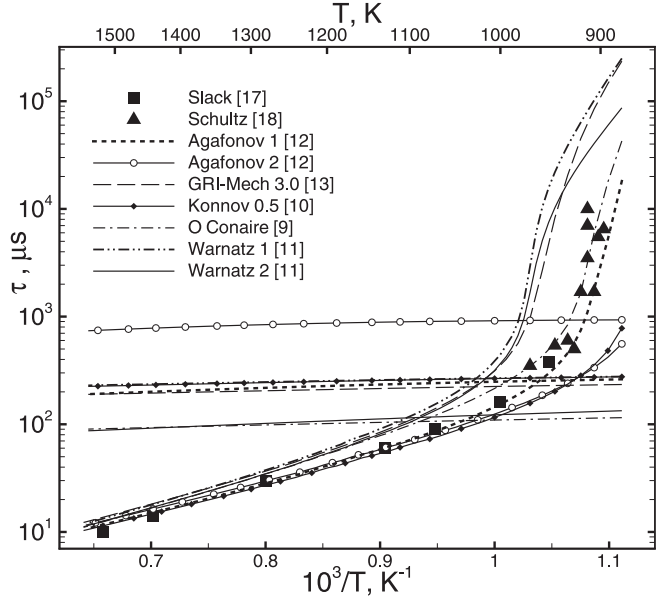}
\caption {\label{Fig1}
Induction periods and exothermal stage durations dependencies
on initial temperature of  stoichiometric hydrogen-air mixture
at 1atm. Experimental data are from \cite{Slack1977,Shepherd2000}.}
\end{figure}

Figure~\ref{Fig1}  shows induction periods and exothermal stage durations
dependencies on initial temperature for stoichiometric
hydrogen-air mixture at 1atm calculated using different
chemical schemes  \cite{OConaire2004,Konnov2004,Warnatz2006,Agafonov1998,GRI}
One can observe almost linear curves intersecting the induction
period dependencies - these curves represent an exothermal
stage duration which determines the time scale of energy
release inside the reaction zone.

%%%%%%%%%%%%%%%%%%%%%%%%

\begin{figure}
\vspace*{1mm} \centering
\includegraphics[width=9cm]{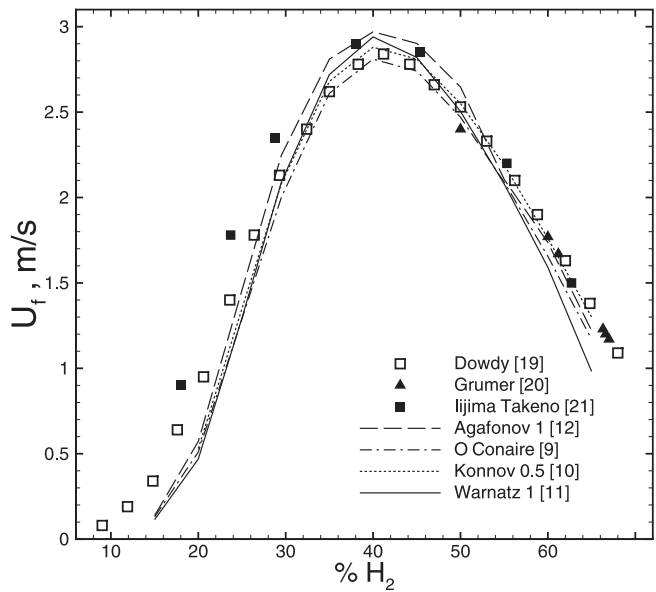}
\caption {\label{Fig2}
Laminar flame speed dependence on equivalent ratio of hydrogen-air
mixture. Experimental data are from \cite{Dowdy1991,Grumer1995,Iijima1986}.}
\end{figure}

We analyzed here 7 schemes by Agafonov \cite{Agafonov1998},
GRI group \cite{GRI}, Konnov \cite{Konnov2004}, O Conaire \cite{OConaire2004}
and Warnatz \cite{Warnatz2006}. The schemes \cite{Agafonov1998}
and \cite{Warnatz2006} are presented in two variants with
different models for three-body collisions.
This factor gives negligible difference in the induction
periods at normal pressure but may affect the time scales
of energy release as it can be seen from Fig.~\ref{Fig1}.
Despite rather sensible differences in the low temperature
region almost all the kinetic schemes reproduce quite close
values of the induction periods at normal conditions.
Among the analyzed schemes one can clearly  extract three
main groups of kinetic schemes reproducing almost the same
values of the induction periods: 1) Agafonov-1 and Konnov,
2) Agafonov-2 and O Conaire, 3) two variants of Warnatz schemes
and GRI scheme. The better agreement with an experimental data
belongs to the second group together with quite sensible differences
between members of the group. Further we will use only four
schemes: one from the first group, one from the third,  one and two
from the second one. The calculations of the laminar flame
speeds $U_f$ at normal pressure (1atm) for different equivalent
ratio of hydrogen-air mixture for four schemes are
presented in Figure~\ref{Fig2}.

The parameters of laminar flame for stoichiometric hydrogen-air
mixture at different pressures, calculated for kinetic
schemes \cite{OConaire2004,Konnov2004,Warnatz2006,Agafonov1998}
are shown in Table 1 together with available
experimental values (Refs.) and calculations from the original
papers where the kinetic schemes were presented and validated
by their authors. One can see that all these chemical schemes
give close values for thermodynamic parameters and for induction
times within accuracy $(10\div15)$$\% $. The main differences are
for low temperature values of induction period.
However, in many cases this part is not essential if endothermic
induction time is larger than characteristic gasdynamic
time of the problem.

\begin{figure}
\vspace*{1mm} \centering
\includegraphics[width=9cm]{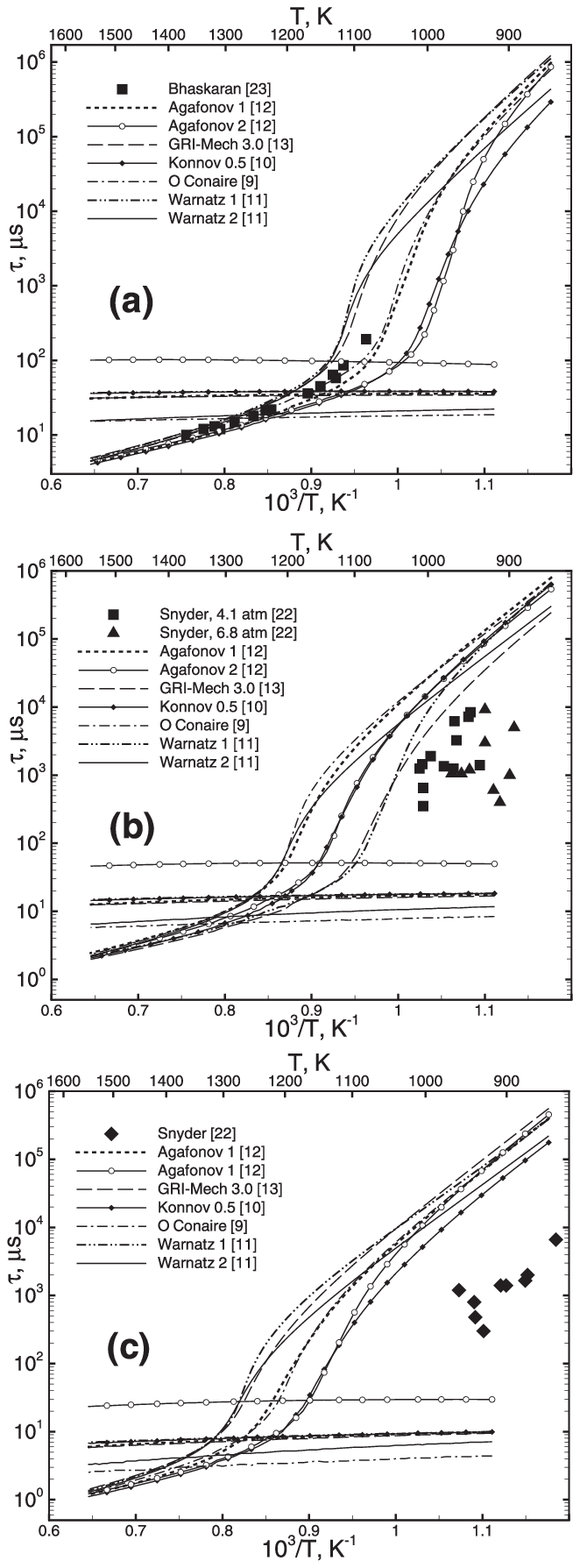}
\caption {\label{Fig3} Induction periods and exothermal stage
durations dependencies on initial temperature of
stoichiometric hydrogen-air mixture at
2.5atm (a), 5.0atm (b) and 8.8atm(c). Experimental
data are from \cite{Snyder1965,Bhaskaran1973}.}
\end{figure}

More essential difference appears for induction time calculated
using different kinetic schemes at elevated pressures.
The induction time calculated using different schemes
at $P=$ 2.5, 5, and 8.8atm are shown in Figures~\ref{Fig3}.
The corresponding parameters of laminar flame are presented in
the Table 1 together with the data for 1atm.
It is seen from Fig.~\ref{Fig3} that considerable difference from
the experimental values for the induction period emerges
in the low temperature region at pressure
greater then 2.5atm as three-body collisions are essential.

\begin{figure}
\vspace*{1mm} \centering
\includegraphics[width=9cm]{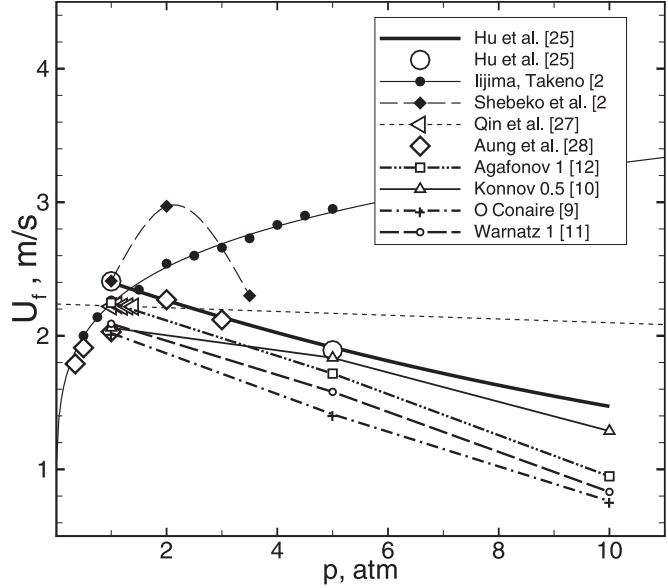}
\caption {\label{Fig4}
Laminar flame speed-pressure dependence for stoichiometric
hydrogen-air mixture. Experimental data are
from \cite{Iijima1986,Hu2009,Shebeko1995,Qin2000}.}
\end{figure}

From Table 1 it is seen that the distinctions in laminar
flame speed also rise with the pressure. The velocity-pressure
dependence calculated using different kinetic schemes and
experimental data together with analytical correlations are
presented in Figure~\ref{Fig4}.

\begin{table}
\label{tab1}
\begin{tabular}{|l|c|c|c|c|c|c|c|c|c|}
\multicolumn{8}{c}{Table 1}\\
\multicolumn{8}{c}{Parameters of laminar $H_2$-air flame given by
the  schemes ([9-12] at $P=1,5,10$bar} \\
\hline
Scheme & $U_f$,m/s & $L_f$,mm & $\Theta$ & $T_b$,K & $H_2$ & ${H_2}O$ & H & $Le$
\\
\hline
$P$(bar) &$1$ &$1$ &$1$ &$1$ &$1$ &$1$ &$1$ & $1$
\\
\hline
OCon & 2.02  (2.00  [9]) & 0.437 & 6.19 & 2168.7 & 0.0094 & 0.3328 & 0.00082 & 1.197
\\
\hline
Kon 0.5 & 2.06  (2.03 [22]) & 0.462 & 6.07 & 2134.9 & 0.0131 & 0.3271 & 0.00168 & 1.193 \\
\hline
Warn 1 & 2.09  (2.00 [23]) & 0.398 & 6.05 & 2128.9 & 0.0134 & 0.3267 & 0.00176 & 1.193
\\
\hline
Agaf 1 & 2.25   & 0.405  & 6.06 & 2132.6 & 0.0160 & 0.3233 &0.00213 & 1.192
\\
\hline
Ref.  & 2.13 [15]  2.35 [24] & - & - & 2138 [11] & 0.017 [11]& 0.320 [11]& 0.002 [11]& -
\\
\hline
$P$(bar) & $5$   & $5$ & $5$ & $5$ & $5$ & $5$ & $5$ &$5$
\\
\hline
OCon   & 1.41 & 0.099 & 6.26 & 2210.5 & 0.0043 & 0.3400 & 0.00017 & 1.198
\\
\hline
Kon 0.5 & 1.84 & 0.103 & 6.21 & 2198.7 & 0.0047  & 0.3397 & 0.00020 &1.195
\\
\hline
Warn 1 & 1.58  & 0.095 & 6.22 & 2200.2 & 0.0044 & 0.3402 & 0.00017 & 1.195
\\
\hline
Agaf 1 & 1.72 & 0.096  & 6.21 & 2196.8 & 0.0050 & 0.3395 &0.00017 & 1.195
\\
\hline
$P$(bar) & $10$ & $10$ & $10$ & $10$ & $10$ & $10$ & $10$ &$10$
\\
\hline
OCon   & 0.76 & 0.071 & 6.28 & 2215.9 & 0.0036 & 0.3412 & 0.00012 & 1.198
\\
\hline
Kon 0.5 & 1.29 & 0.078 & 6.24 & 2205.7 & 0.0034  & 0.3415 & 0.00011 &1.195
\\
\hline
Warn 1 & 0.83  & 0.072 & 6.26 & 2205.2 & 0.0034 & 0.34015 & 0.00011 & 1.195
\\
\hline
Agaf 1 & 0.95 & 0.074  & 6.25 & 2202.6 & 0.0041 & 0.3408 &0.00011 & 1.195
\\
\hline

\end{tabular}
\end{table}

One can see that some of the correlations obtained from the
analysis of the experimental data are quite distinct from the
obtained calculations and from each other not only quantitatively
but even qualitatively. One of the most recent is the correlation
obtained in \cite{Hu2009}. It agree well with several
experimental data points and shows a qualitative
velocity-pressure dependence close to the obtained
numerically using different schemes.

\section{Hydrogen-oxygen mixture}

\begin{figure}
\vspace*{1mm} \centering
\includegraphics[width=9cm]{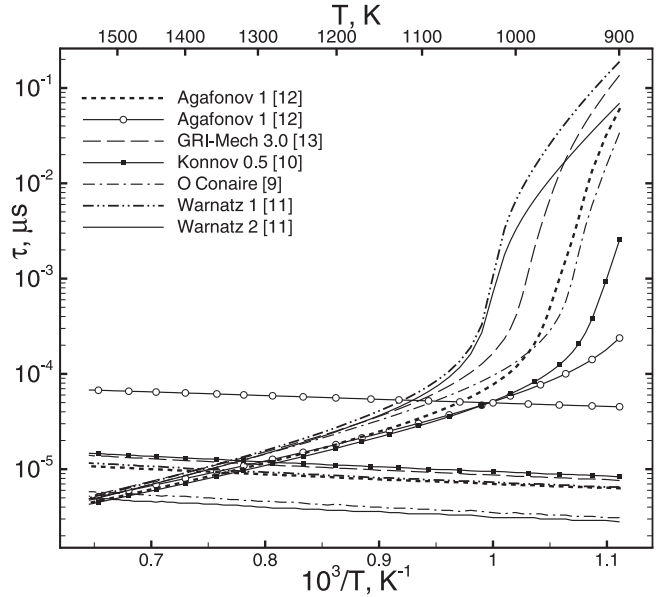}
\caption {\label{Fig5}
Induction periods and exothermal stage durations dependencies
on initial temperature of stoichiometric hydrogen-oxygen mixture at 1atm.}
\end{figure}

The difference in time scales and corresponding difference in
flame width become even more noticeable when it is calculated
using different kinetic schemes for highly reactive mixtures
as e.g. hydrogen-oxygen. Correspondingly, the larger distinction
is found for the flame speeds. Figure~\ref{Fig5}  shows induction
periods and exothermal stage durations dependencies on initial
temperature calculated for stoichiometric
hydrogen-oxygen mixture  at 1atm for different chemical schemes \cite{OConaire2004,Konnov2004,Warnatz2006,Agafonov1998,GRI}.
The main difference in the induction time given by different
schemes is again in the region of low
temperature (see comment to Fig.~\ref{Fig1}).

\begin{figure}
\vspace*{1mm} \centering
\includegraphics[width=9cm]{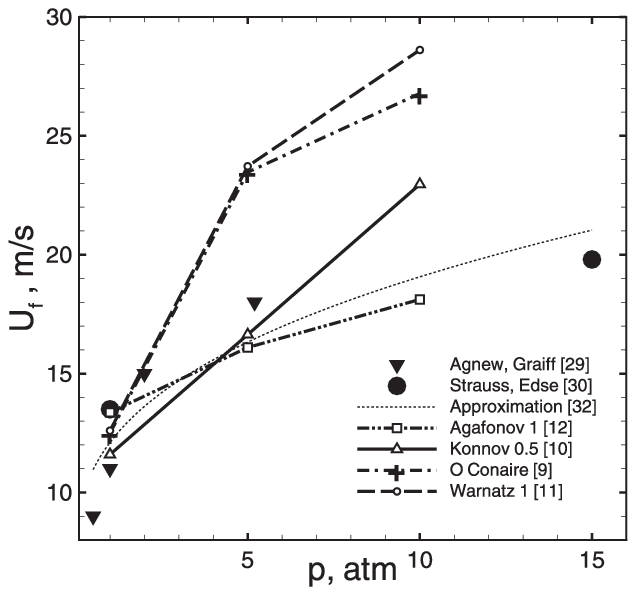}
\caption {\label{Fig6}
Laminar flame speed-pressure dependence for stoichiometric hydrogen-oxygen mixture.
Experimental data is taken from \cite{Agnew1961,Strauss1958}.}
\end{figure}

Figure~\ref{Fig6} shows velocity-pressure dependence of laminar
flame calculated for chemical
schemes \cite{OConaire2004,Konnov2004,Warnatz2006,Agafonov1998}.
It is interesting  to notice that experimental data of the
velocity-pressure dependence can be
approximated as ${U_f} \propto {P^{{n \over 2} - 1}}$  well known
from classical combustion theory \cite{Zeld1985}  with overall
reaction order 2.74 presented in \cite{KuzLib2010} .
The approximation is shown by the dotted line in Figure~\ref{Fig6}.
It should be also noted that there are almost no qualitative
difference in the velocity-pressure dependencies in the
hydrogen-oxygen scheme. It may be caused by the constant
behavior of the overall reaction order.
In hydrogen-air mixture the overall reaction order changes
with the pressure and the difference rises with the
nitrogen dilution as it shown in \cite{Egolfopoul1991}.

\section{Conclusions}

The objective of the present study was to evaluate different
reduced chemical kinetic schemes with the purpose to
understand their applicability and reliability for numerical
modeling of the complex multiscale phenomena of unsteady multidimensional
combustion, which is typically characterized by a flow with large
gradients of temperature and pressure. While speed of sound and
therefore characteristic hydrodynamic time scales do not depend
on pressure, the induction time, especially at the temperature
range $(1000 \div 1200)$K is considerably sensitive to pressure.
This and different pressure dependencies given by different reduced schemes
must be taken into account while modeling unsteady combustion processes.
Comparison of different kinetic models and criteria of their agreement
with experimental data for the velocity-pressure dependence and
width of the flame probably proved to be an effective guide for
option of the most reliable kinetic model. To what extend the range
of the model applicability should be considered as trustable
is not certain due to the lack of available experimental data.

\begin{acknowledgements}
The computations were performed on resources provided by
the Swedish National Infrastructure for Computing (SNIC) at
the Center for Parallel Computers at the Royal Institute of
Technology in Stockholm and the National
Supercomputer Centers in Linkoping.
\end{acknowledgements}

\end{document}